# Stabilization of class-B broad-area lasers emission by external optical injection

A.V. P‌AKHOMOV,[1,2,*] R.M. A‌RKHIPOV,[3,4,5] N.E. M‌OLEVICH[1,2]

[1]*Department of Physics, Samara University, Moskovskoye Shosse 34, Samara 443086, Russia*
[2]*Department of Theoretical Physics, Lebedev Physical Institute, Novo-Sadovaya Str. 221, Samara 443011, Russia*
[3]*Max Planck Institute for the Science of Light, Staudtstraße 2, 91058 Erlangen, Germany*
[4]*St. Petersburg State University, Faculty of Physics, Ulyanovskaya Str. 3, St. Petersburg 198504, Russia*
[5]*ITMO University, Kronverkskiy prospekt 49, St. Petersburg 197101, Russia*
*\*Corresponding author: antpakhom@gmail.com*

We theoretically examine the effect of external optical injection on the spatio-temporal dynamics of class-B broad-area lasers. We demonstrate that optical injection can efficiently stabilize the intrinsic transverse instabilities in such lasers associated with both the boundaries of the pumping area and with the bulk nonlinearities of the active medium. Stabilizing action of optical injection is shown to be closely related to the suppression of inherent relaxation oscillations behavior.

http://dx.doi.org/10.1364/AO.99.099999

## 1. INTRODUCTION

Development of ultra-compact light sources with stabilized parameters is in great demand in modern optics and gains considerable technological importance. These devices can be useful in different applications such as data processing and optical transmission systems. Compact laser sizes also allow to easily achieve single-frequency operation and fast response time due to short cavity length. However, small cavity length results in sufficiently low output power what constraints laser applicability to only short-range optical data transmission. The most natural way to circumvent this limitation and to increase the output power without rejection of all advantages of short-cavity configuration is the usage of the broad-area lasers with wide transverse section. This approach is relevant in particular for microchip solid-state lasers or vertical-cavity surface-emitting semiconductor lasers. The price for the corresponding benefit is the deterioration of characteristics of the emitted beam due to arising competition between transverse spatial modes that can be important for even small-area lasers [1] and plays ever increasing role as long as the number of excited modes grows.

Transverse laser instabilities are responsible for filamentary dynamics in the cross-section of the emitted beam leading to sufficient reduction of its spatial and temporal coherence. Though in some applications like projection systems low spatial coherence of laser emission may be even extremely desirable as it allows reducing the speckle contrast [2], usually these complex regimes are assumed to be a harmful effect. Therefore for practical reasons it is highly desirable to suppress these spatio-temporal instabilities.

The complex spatio-temporal dynamics owing to arising transverse instabilities in free-running broad-area lasers was well established in a row of experimental studies [3,4]. Filamentary dynamics usually originates from intrinsic self-focusing nonlinearity conditioned with either strong phase-amplitude coupling typical for semiconductor active mediums [5,6] or with Kerr-type material nonlinearity [7].

However, even in the absence of these specific self-focusing mechanisms filamentary instabilities can arise. This may be provided due to either the bulk medium instabilities [8-10] or the effects on the edges of the pumping profile [10-14]. In the former case stabilization of the lasing profile was observed for some specific laser parameters in [15] by limiting the transverse extension of pumping to be about the scale of the unstable field components so that these unstable modes could not develop. The influence of the boundaries of the pumping region was found to be suppressed in the case of strongly inhomogeneous broadening of the gain profile of the active medium [16-18].

The case of the lasers of class-B dynamical limit is of special interest, meaning the coherence relaxation in the active laser medium is very fast and the population inversion relaxation is very slow as compared to the photon decay rate in the cavity. This class of lasers is of greatest importance for application, as it includes most of the high-demand laser types (solid-state, semiconductors). Class-B lasers exhibit especially complex spatio-temporal dynamics associated with the competition of either longitudinal modes [19-21] or transverse modes in small-area lasers [22-25]. It was also found that in broad-area class-B lasers filamentary instabilities become especially well-pronounced. Specifically, class-B broad-area lasers were shown to exhibit two main types of intrinsic transverse instability mechanisms leading to the filamentary dynamics of the laser output [10]. Firstly, the homogeneous output profile turns to be intrinsically unstable in class-B limit due to the bulk instability even in the absence of the phase-amplitude coupling in the active medium and independently from the boundary conditions. Secondly, high sensitivity to the boundaries of the pumping region was found for the commonly used top-hat-like profile leading to boundaries-induced filamentary dynamics. The harmful action of these effects encourages finding effective approaches for the suppression of the transverse filamentary instabilities.

Recently a new method for the suppression of self-focusing instability in semiconductor amplifiers and lasers was elaborated that is based on the spatially periodic modulation of the pump profile in both longitudinal and transverse directions [26-29]. This stabilizing effect was found to take place in both the linear stage of the amplification and in the nonlinear regime when the modulation instability arises. Later a similar method was examined for vertical-cavity semiconductor lasers with the spatio-temporal modulation of the pumping profile both in time and in two transverse spatial dimensions [30,31]. This stabilization method was shown to operate most efficiently in class-A laser limit (for relatively long VECSEL resonators), while it becomes ineffective in class-B laser limit (for relatively short resonators). In paper [32] another alternative approach was proposed that relies on the use of the specially designed intracavity photonic crystals introducing the refractive index modulation in both the longitudinal and the transverse dimensions. A substantial improvement of the spatial quality of the output beam emitted by microchip laser was then obtained due to the spatial filtering functionality of intracavity photonic crystals.

In this paper we study the influence of external optical injection on the spatio-temporal dynamics of class-B broad-area laser. Injection of the external optical signal is the well-known method to enrich the laser dynamics allowing formation of the cavity solitons under bistability conditions [33,34] or even more complex regimes [35]. Besides that, optical injection is also the well-established way to stabilize the parameters of lasers emission. In particular, it has been demonstrated experimentally and theoretically that locking the slave laser to the master one allows to reduce the noise-induced spectral broadening and stabilize the pulse waveform and leads to reduction of the timing jitter in mode-locked semiconductor lasers [36-40]. Strong optical injection was also found to overcome the self-focusing action owing to the gain-index coupling and suppress filamentation in broad-area semiconductor lasers [41]. The effect of the external injection on the intrinsic class-B filamentary instabilities does not seem to have been studied until now. We show in this paper that weak coherent injection can act as an efficient way to completely eliminate these transverse instabilities in class-B broad-area lasers.

The paper is organized as follows. In Section 2 we present the theoretical model of an optically-injected broad-area laser used in our analysis. In Section 3 we describe the stabilizing impact of the optical injection on the boundaries-induced filamentation onset together with the corresponding transient dynamics. Section 4 is devoted to the bulk transverse instabilities in class-B broad-area lasers. We performed the stability analysis of the stationary lasing state and found out that unstable perturbation modes occurring in the free-running laser would be suppressed when laser is submitted to optical injection. Finally, a conclusion is presented. Additionally, derivation of the analytical results is given in the Appendix.

## 2. THEORETICAL MODEL

Our analysis is based on Maxwell-Bloch equations for broad-area optically-injected laser with homogeneously broadened two-level media inside Fabry-Perot cavity and in the mean-field approximation [42] (schematic representation is provided by Fig. 1):

$$\begin{cases} \dfrac{\partial E}{\partial t} = \kappa \left[ P - E(1 - i\delta) + E_{inj}\, e^{i\theta t} \right] + i\,\tilde{a}\,\Delta_\perp E, \\ \dfrac{\partial P}{\partial t} = \gamma_\perp \left[ -(1 + i\delta)P + DE \right], \\ \dfrac{\partial D}{\partial t} = -\gamma_\parallel \left[ D - r(x,y) + \dfrac{1}{2}\left( E^* P + E P^* \right) \right], \end{cases} \quad (1)$$

where $E$, $P$, $D$ stand for the dimensionless envelopes of the intracavity electric field, active media polarization and population inversion, respectively; $\gamma_\perp$, $\gamma_\parallel$ and $\kappa$ are the coherence relaxation rate, population inversion relaxation rate and electric field decay rate, respectively. It is convenient to rescale time in the system (1) to the coherence lifetime $\gamma_\perp^{-1}$ with introducing the timescales ratios $\gamma = \gamma_\parallel / \gamma_\perp$ and $\sigma = \kappa / \gamma_\perp$. In presented notation, the class-B laser condition is given as: $\gamma \ll \sigma \ll 1$.

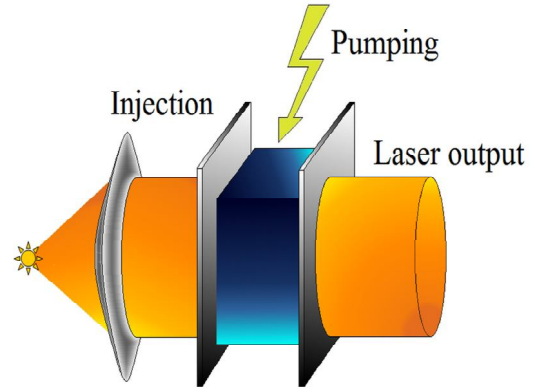

Fig. 1. Schematic representation of an optically-injected broad-area laser.

Spatial coupling in the system (1) is provided by the light diffraction which is taken into account through the Laplacian term in the field equation. We suppose the transverse spatial coordinates $x$ and $y$ normalized to the characteristic transverse spatial scale of the laser $d$; normalized diffraction coefficient is then given as $a = c^2 / \left( 2\omega_{\text{cav}} \gamma_\perp n_{gr} n_{ph} d^2 \right)$, where $n_{ph}$ and $n_{gr}$ are the phase and group refractive indexes of the host medium. Parameter $r(x,y)$ represents the pumping rate normalized versus its threshold value. $\delta = (\omega_{21} - \omega_{\text{cav}})/(\gamma_\perp + k)$ is the rescaled detuning between the optical transition frequency $\omega_{21}$ and the cavity resonant frequency $\omega_{cav}$. External injection is characterized by two parameters: $E_{inj}$ denotes the injection strength and $\theta$ corresponds to the frequency mismatch for the injected emission. The injection coupling coefficient is proportional to the cavity mirror transmittance and is thus of the same order of magnitude as the cavity decay rate; without loss of generality we consider them to be equal.

The model of two-level active medium is the relevant one to describe correctly the light-matter interaction when taking into account only a single optical resonance with the homogeneously broadened line. This is reasonably justified for the case of the interaction through a separate and well resolved optical transition, especially in the mediums having atomic-like energy-level structure. Such approximation is usually efficient for modeling of the rare-earth active ions in solid-state gain medium or the radiative transitions in molecular and atomic gas lasers. Also, it can be expected to be applicable for low-dimensional semiconductor structures, like quantum dots and dashes, because of their anticipated symmetric shape of the gain spectrum should result in a very weak phase–amplitude coupling. We neglect also the diffusion spatial coupling in the active medium as this process in aforementioned materials is either negligible (doped crystals) or strongly reduced due to carrier confinement (quantum dots and dashes) and thus mostly very slow being compared with the diffraction.

In broad-area lasers the pumping is naturally provided over the large area in the transverse section. The spatial profile of the pumping rate is therefore commonly reasoned to be slowly varying over the whole area where the population inversion is maintained. For this purpose top-hat profiles are widely used that are close to uniform in the central part and slowly decrease towards the profile edges. To model the finite-size pumping area with smoothed top-hat-like cross section we take here the Fermi-Dirac-type profile in the form:

$$r(x,y) = r_0 \left( \frac{2}{\left(1+e^{\frac{|x-x_0|-L}{D}}\right)\left(1+e^{\frac{|y-y_0|-L}{D}}\right)} - 1 \right). \quad (2)$$

Negative values of the pumping outside the central core describe noninversed absorbing medium in the unpumped regions. Parameters $L$ and $D$ in (2) describe, respectively, the effective width of pumping region and the profile slope near the edges. Varying these parameters allows to control the profile shape in the wide limits and thus the onset conditions of the transverse instabilities. Under external injection it may be expected that the pumping shape would be of great importance.

Class-B lasers are known to exhibit extremely unstable spatio-temporal behavior resulting in filamentary operation even in the absence of the gain-index coupling or other explicit self-focusing nonlinearity. Specifically, with widely used top-hat pumping profile output emission turns to be intrinsically driven to the filamentary state due to the high sensitivity to boundaries [10-14]. We examine the impact of external optical injection on these effects in the next section.

## 3. STABILIZING EFFECT OF OPTICAL INJECTION

It seems to be the most practically relevant to achieve the stable emission for as low injection strength as possible. For the external field that is not coherent with the laser under study, the injection must be strong enough to lock the laser. Therefore we restrict ourselves in the following to the case of coherent injection, that is $\theta = 0$. For coherent injection its frequency should be equal to the laser emission frequency that is given by the mode pulling formula:

$$\omega_0 = \frac{\omega_{12} + \sigma\omega_{cav}}{\sigma + 1}.$$

Thus we stay the complex dynamics observed outside the locking parameter region [43] out of the scope of this paper. We expect, however, the obtained results to stay valid for incoherent injection as well, but at larger values of injection strength.

For proof-of-principle investigation we would neglect first the injection wavefront curvature thus considering it being close to plane within the pumped region. We report here that injected field can have a profound stabilizing effect for even relatively weak injection. To be more exact, we consider the case of such optical injection that the amplitude of the injected field is small compared to the amplitude of the intracavity laser field, i.e. $|E_{inj}| << |E|$. The latter restriction corresponds to the laser cavity where injected field does not play a driving role in laser operation. It may seem to be more reasonable to compare the injection field amplitude to the laser output field, that is $T \cdot |E|$ with $T$ being the cavity output mirror transmissivity. However, the commonly encountered values of mirror transmissivity can vary over wide range, for which reason we will proceed in the following to be definite from the former criterion.

We fix for definiteness the following parameter values identical to those previously used in [10]: $\gamma_\perp^{-1}$ = 50 fs, $\gamma_\parallel^{-1}$ = 1 ns, $\kappa^{-1}$ = 2 ps, $\omega_{cav}$ = 2.5·10$^{15}$ s$^{-1}$, $n_{ph}$ = 3.6, $n_{gr}$ = 4.2, $d$ = 10 $\mu$m. Such parameters values are typical for relaxation times, optical coefficients and characteristic transverse scales in semiconductor lasers; that yields for the dimensionless parameters the following values: $\sigma$=0.025, $\gamma$=5·10$^{-5}$ and $a$=6·10$^{-4}$. Pumping level $r_0$ and frequency detuning $\delta$ were considered to be the control parameters.

Performed numerical simulations showed that increasing the injection value results in decrease of the transverse intensity modulation leading to the filamentary output pattern fading out. An example is shown in Fig. 2 for fixed parameter values $r_0$=2, $D$=0.7, $L$=7, $\delta$=0 and different values of the injection strength. Importantly, when exceeding some definite injection threshold the transverse modulation becomes completely suppressed and the highly-coherent nonfilamented output beam with top-hat profile is attained. Obtained flat-top beam shape naturally follows from the pumping profile and seems to be of particular importance since such beams are practically relevant for many applications.

The stabilization threshold was found to sensitively depend on the pumping profile shape. Specifically, with smoothing the sharp edges of the pumping profile the threshold value decreases displaying reduced sensitivity to the boundaries effects. Obtained dependence is illustrated by Fig. 3, where the total threshold reduction by about a factor of five is found within the considered range of the edge slope parameter in Eq. (2). It is important to notice that, as can be seen from Fig. 2, increase of the injection strength gives rise to the increase of the transient processes damping. Namely, strong suppression of the intrinsic class-B relaxation oscillations is observed together with the disappearance of the spiking behavior in the early stages of the transient response.

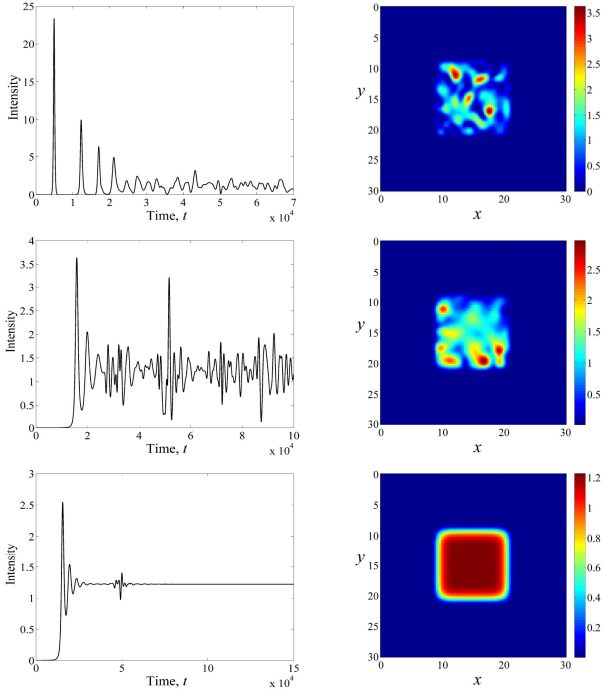

Fig. 2. Laser output dynamics for $r_0$=2, $\delta$=0, $D$=0.7, $L$=7; left column - intensity time series; right column - transverse intensity profile. $|E_{inj}|$=0 (top line), 0.01 (middle line), 0.02 (bottom line).

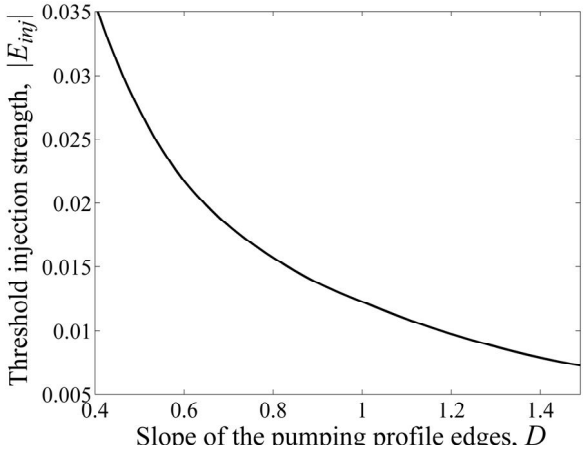

Fig. 3. The dependence of the threshold injection amplitude on the pumping profile shape; $L$=7, $r_0$=2, $\delta$=0.

We turn now to the more realistic case of the Gaussian shape of the injection beam:

$$E_{inj} = E_{inj}^0 \, e^{-\frac{x^2+y^2}{w_0^2}}, \quad (3)$$

where $E_{inj}^0$ is the injected field amplitude and $w_0$ is the beam waist.

Fig. 4 shows the obtained dependence of the threshold injection strength on the waist of injection beam (3). As could be expected for the boundaries-induced instability, the stabilization injection threshold depends heavily on the injection beam amplitude near the pumping edges. This dependence appears to be especially strong when the beam waist is close to the width of the pumping region. With further increase of the injection beam waist the threshold injection strength rapidly decreases and tends to the limiting value corresponding to the plane injection wavefront.

## 4. SUPPRESSION OF BULK INSTABILITIES

Stabilization of boundaries-induced onset of filamentation found in the previous section can not, however, guarantee the nonfilamented output in real-world broad-area lasers since the bulk instability mechanisms independent of boundaries effects may come into play. In considered Maxwell-Bloch equations such inherent instabilities are especially well-pronounced in class-B case resulting in filamentary dynamics even in boundaries-free cavity [10]. The solitary class-B lasers exhibit the traveling-wave instability that emerges at nonzero, but even very small values of frequency detuning and initiates the filamentary behavior of the laser output. In real-world lasers the presence of the frequency detuning can be naturally conditioned by the thermal heating of the active medium. Hence, it turns out to be crucial to study the influence of optical injection on this bulk transverse instability.

The starting point for our analysis is the model (1), where we set aside now the shape of the pumping profile and consider it to be spatially uniform. Steady states $E_{st}$ of the corresponding boundaries-free problem are given as:

$$E_{inj} = E_{st}\left(1 - i\left(\delta - \frac{\theta}{\sigma}\right)\right) - \frac{rE_{st}(1-i(\theta+\delta))}{|E_{st}|^2 + 1 + (\theta+\delta)^2}. \quad (4)$$

For coherent injection equation (4) has the unique feasible solution above the lasing threshold. Stability analysis for this solution results in rather cumbersome characteristic equation and is given in the Appendix section.

According to Eq. (A2) at a first approximation we obtain the following additional injection-induced term in the expression for the growth rate of the unstable branch of characteristic equation:

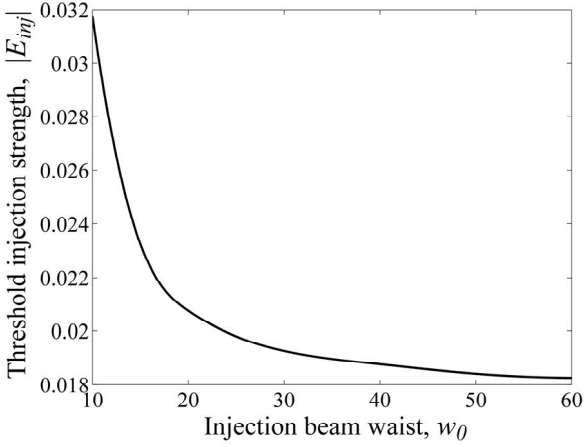

Fig. 4. The dependence of the threshold injection strength on the Gaussian injection beam waist; $L=7$, $D=0.7$, $r_0=2$, $\delta=0$.

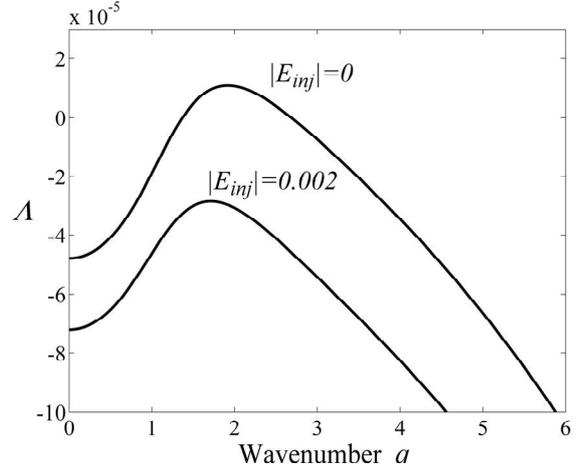

Fig. 5. An example of the dispersion curves for different values of the injection strength; $r_0=2$, $\delta=-0.1$.

$$\Lambda^{inj}(q) = -\frac{\sigma(a^2 q^4 + (1-\delta^2)\sigma\gamma I_{st})}{(1+\delta^2)(2\sigma\gamma I_{st} + a^2 q^4(1+\delta^2))} \times \left(\frac{E_{inj}(1+i\delta)}{E_{st}}\right). \quad (5)$$

The lattermost multiplier in Eq. (5) is positive as it follows from Eq. (4). Considering the practically relevant values for the frequency detuning $\delta$ are no more than about several tenths, the additive component $\Lambda^{inj}(q)$ (5) is negative for all values of the wavenumber $q$ and tends to constant values as $q \to 0$ and $q \to \infty$. Eq. (5) hence describes the downward translation of the whole dispersion curve that is parallel translation for small and large values of $q$ and entails certain change in curve shape for the moderate values containing the unstable modes. As the dispersion curve displacement is proportional to the injection strength, the complete suppression of instability should take place as long as the injection exceeds a certain threshold.

Numerical stability analysis confirmed aforesaid tendency for even larger injection strength. Importantly, an application of optical injection was found to result in downward translation of the whole dispersion curve leading to elimination of unstable field components for all the considered values of the control laser parameters $\delta$, $r$. Fig. 5 shows an example of the dispersion curves containing unstable wavenumbers for the solitary laser and absolutely stable when laser is subject to coherent optical injection. Suppression of the bulk instability in this case has the threshold nature as well.

Corresponding injection threshold values were obtained to be sufficiently less than those for boundaries-induced filamentary instability. Fig. 6 illustrates the dependence of the threshold injection amplitude on the control parameters. It is well seen that for moderate distances from the lasing threshold the weak injection condition is fulfilled by a wide margin. Obtained stability results mean the optical injection to stabilize the broad-area laser output for both types of filamentary origin thus providing the nonfilamented output regardless of what instability mechanism is the prevailing one in each particular case.

Stabilizing action of optical injection seems to be closely related to the suppression of inherent relaxation oscillations behavior. Indeed, Eq. (A1), when assuming $q=0$, allows to obtain explicitly the expressions for the transients damping rates in the class-B limit $\gamma \ll \sigma \ll 1$ (see Eq. (A4)):

$$\Gamma_{RO} = \frac{\gamma r}{2(1+\delta^2)} + E_{inj}\frac{\sigma(1+i\delta)(1-\delta^2)}{2E_{st}(1+\delta^2)} + \dots; \quad (6)$$

$$\Gamma_{ADD} = 0 + E_{inj}\frac{\sigma(1+i\delta)}{E_{st}} + \dots. \quad (7)$$

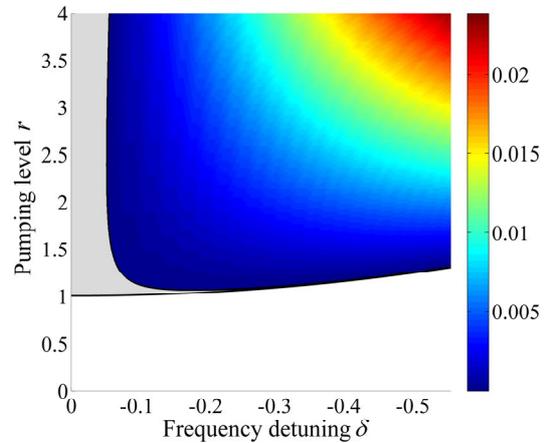

Fig. 6. Threshold injection amplitude $|E_{inj}|$ vs the pumping level $r$ and the frequency detuning $\delta$; lower solid curve describes the lasing threshold; single-color gray area on the left side of the figure above the threshold corresponds to the stability region.

Parameter $\Gamma_{RO}$ (6) having nonzero value in the absence of injection corresponds to the conventional damping rate of the relaxation oscillations. $\Gamma_{ADD}$ (7) gives additional damping rate which appearance is related to the breaking of the phase invariance in optically-injected case. With practically relevant values for frequency detuning $\delta$ not exceeding several tenths, both damping rates monotonically increase with the increase of injection amplitude. This dependence is illustrated in Fig. 7. The actual damping of the transient processes is governed by the lowest of both damping rates. For small detuning values additional damping rate $\Gamma_{ADD}$ increases approximately twice as rapidly with injection as $\Gamma_{RO}$, that is why their relative contributions to the actual transient response vary with the injection strength.

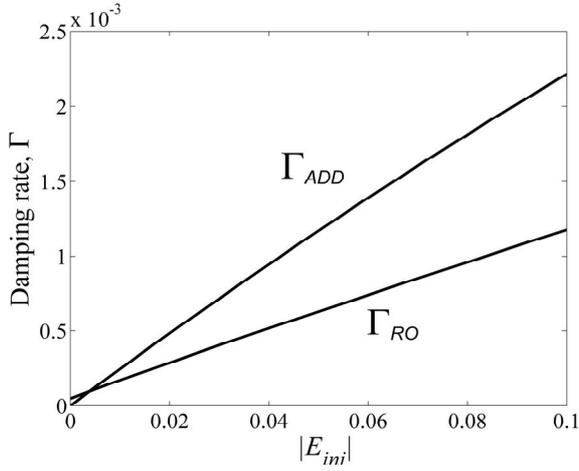

Fig. 7. Damping rates for the transient processes against the injection amplitude; $r_0=2$, $\delta=-0.1$.

Obtained dependence shows that the relaxation oscillations may be strongly suppressed in a laser submitted to optical injection. Since the relaxation oscillations are the key property of class-B lasers and play a decisive role in their dynamics, it seems to be happening that external injection modifies the overall laser dynamical features. We thus tend to think the obtained stabilization of inherent class-B laser transverse instabilities to originate from the forced suppression of the intrinsic relaxation oscillations dynamics.

## 5. CONCLUSIONS

We theoretically demonstrated that optical injection can enable efficient stabilization of intrinsic transverse instabilities in class-B broad-area lasers. Based on Maxwell-Bloch equations, we considered in detail the influence of injected emission on the main instability mechanisms, namely associated with the effect of the pumping region boundaries and with the bulk nonlinearities of the active medium.

The boundaries-induced onset of filamentary dynamics was found to be suppressed in the presence of injected signal. As the result, the highly-coherent beam with flat-top profile and high beam quality was obtained at the laser output. Notably, this effect appears to be of threshold nature, so that the laser exhibits some irregular transverse intensity modulation diminishing with the increase of injection strength and vanishing at the threshold injection value. Revealed stabilization turns out to be sufficiently sensitive to the pump profiling with increasingly enhanced efficiency as long as the profile shape becomes more smoothed.

Similar effect was found for the onset of the traveling-wave instability related to the bulk medium nonlinearities. We demonstrated that injection efficiently suppresses the unstable field components thus allowing for the stable output beam regardless of boundaries influence. The stabilization of bulk instability shows better results compared to the boundaries-controlled case. Specifically, for low or moderate distances from the lasing threshold the injection amplitude required for the complete instability suppression proves to be strongly reduced.

Besides the transverse instabilities, optically-injected laser exhibits improved time-domain performance arising from the stronger damped transient response. Therefore we attribute the revealed laser output stabilization to the injection-induced suppression of inherent relaxation oscillations behavior.

## Appendix

The characteristic equation for the stability analysis of an optically-injected laser can be represented in the form:

$$\lambda^5 + a_1^0 \lambda^4 + (a_2^0 + a_2^{inj})\lambda^3 + (a_3^0 + a_3^{inj})\lambda^2 + (a_4^0 + a_4^{inj})\lambda + (a_5^0 + a_5^{inj}) = 0, \quad \text{(A1)}$$

where coefficients with "0" subscript correspond to the free-running laser in the absence of injection:

$$a_1^0 = 2(1+\sigma) + \gamma;$$

$$a_2^0 = 1 + \delta^2 + 2\sigma + \sigma^2 - 2\sigma\delta^2 - 2\sigma\delta aq^2 + a^2 q^4 + \sigma^2\delta^2 + \gamma(2 + 2\sigma + I_{st});$$

$$a_3^0 = -4\sigma\delta aq^2 + 2a^2 q^4 + \gamma\big(r + 2\sigma + \sigma^2 + 3\sigma I_{st} + a^2 q^4 - 2\sigma\delta^2 + \sigma^2\delta^2 - 2\sigma\delta aq^2\big);$$

$$a_4^0 = a^2 q^4 (1+\delta^2) + \gamma\big(2\sigma I_{st} + 2\sigma^2 I_{st} + 2a^2 q^4 + a^2 q^4 I_{st} - 4\sigma\delta aq^2 - \sigma\delta I_{st} aq^2\big);$$

$$a_5^0 = \gamma\big(ra^2 q^4 - 2aq^2 \delta\sigma I_{st}\big),$$

and coefficients with subscript "inj" are related to injection:

$$a_2^{inj} = \frac{2\sigma E_{inj}(1+i\delta)}{E_{st}};$$

$$a_3^{inj} = \frac{2\sigma E_{inj}(1+i\delta)}{E_{st}}(\gamma+\sigma+1) - \frac{\sigma\gamma}{2}\big(E_{st}^* E_{inj} + E_{st} E_{inj}^*\big);$$

$$a_4^{inj} = \frac{2\sigma E_{inj}(1+i\delta)}{E_{st}}\big(\gamma(1+\sigma+0.5\cdot I_{st}) - \delta aq^2\big) + \left(\frac{\sigma E_{inj}(1+i\delta)}{E_{st}}\right)^2 - \frac{\sigma\gamma}{2}\big(E_{st}^* E_{inj}(\sigma(1-i\delta) + iaq^2 + 1 + i\delta) + E_{st} E_{inj}^*(\sigma(1+i\delta) - iaq^2 + 1 - i\delta)\big);$$

$$a_5^{inj} = \frac{\sigma\gamma E_{inj}(1+i\delta)}{E_{st}}\left(2\sigma I_{st} - 2\delta a q^2 + \frac{\sigma E_{inj}(1+i\delta)}{E_{st}}\right) - \frac{\sigma\gamma}{2}\left(E_{st}^* E_{inj} + E_{st} E_{inj}^*\right)\left(\frac{\sigma E_{inj}(1+i\delta)}{E_{st}}\right),$$

where $I_{st} = |E_{st}|^2$.

To obtain the analytical expressions for the parameters of unstable modes, we will follow the small-parameter expansion procedure proposed in [10]. Since considering the class-B laser limit, we assume the rescaled cavity decay rate $\sigma$ as the order of magnitude for the small parameter $\varepsilon$, that is $\sigma \sim \varepsilon$. For class-B case we take for the normalized inversion relaxation rate $\gamma \sim \varepsilon^3$. Parameters $\delta$, $r-1$ are supposed to take on arbitrary values, but we limit their product to be $\delta I_{st} = \delta(r-1-\delta^2) \sim \varepsilon$ due to its nonmonotonic dependence on $\delta$. We also take for the characteristic parameters of the unstable modes the next estimates [10]: for the unstable wavenumbers $aq_{unst}^2 \sim \varepsilon^2$ and for the unknown characteristic roots $\lambda = \Lambda \pm i\Omega$ we assume $\Lambda \sim \varepsilon^3$, $\Omega \sim \varepsilon^2$.

It can be then obtained from Eq. (A1) that for injection strength above the second order of $\varepsilon$, all terms with injection will be omitted in characteristic equation (A1). That is why we suppose the following estimate for the injection amplitude: $E_{inj} \sim \varepsilon^2$.

With allowance for these assumptions and splitting real and imaginary parts in (A1), we get the expressions for the unstable branch of the dispersion curve at the lowest order of magnitude:

$$\lambda_{1,2} = \Lambda \pm i\Omega,$$

where the imaginary part of the characteristic roots is:

$$\Omega^2(q) = a^2 q^4 + \frac{2\gamma\sigma I_{st}}{1+\delta^2},$$

and the growth rate of the perturbation components:

$$\Lambda(q) = \Lambda^0(q) + \Lambda^{inj}(q)$$

with

$$\Lambda^0(q) = \frac{1}{1+\delta^2}\left(2\sigma\delta a q^2 - \frac{\gamma r}{2}\right) + \frac{\gamma}{2} \times \frac{ra^2 q^4 - 2aq^2\sigma\delta I_{st}}{a^2 q^4(1+\delta^2) + 2\gamma\sigma I_{st}};$$

$$\Lambda^{inj}(q) = -\frac{\sigma\left(a^2 q^4 + (1-\delta^2)\sigma\gamma I_{st}\right)}{(1+\delta^2)(2\sigma\gamma I_{st} + a^2 q^4(1+\delta^2))} \times \left(\frac{E_{inj}(1+i\delta)}{E_{st}}\right).$$ **(A2)**

By taking $\Omega = 0$ we obtain for another characteristic root that appears to be purely real:

$$\lambda_3(q) = -\gamma\,\frac{ra^2 q^4 - 2aq^2\sigma\delta I_{st}}{a^2 q^4(1+\delta^2) + 2\gamma\sigma I_{st}} - \frac{2\gamma\sigma^2 I_{st}}{a^2 q^4(1+\delta^2) + 2\gamma\sigma I_{st}} \cdot \left(\frac{E_{inj}(1+i\delta)}{E_{st}}\right).$$ **(A3)**

Eqs. (A2), (A3) allow to get the expressions for the damping rates of the transient processes upon setting $q=0$:

$$\Gamma_{RO} = \frac{\gamma r}{2(1+\delta^2)} + E_{inj}\,\frac{\sigma(1+i\delta)(1-\delta^2)}{2 E_{st}(1+\delta^2)} + \ldots;$$

$$\Gamma_{ADD} = 0 + E_{inj}\,\frac{\sigma(1+i\delta)}{E_{st}} + \ldots.\quad\textbf{(A4)}$$

Expressions (A4) are represented as the expansions with respect to the injection strength $E_{inj}$. It is thus seen that the first one corresponds to the usual damping rate of relaxation oscillations in the absence of injection and is therefore labeled as $\Gamma_{RO}$. Another additional damping rate $\Gamma_{ADD}$ equals zero without injection and is related to the phase invariance of the system.

**Funding Information.** Ministry of education and science of Russian Federation (GR 114091840046), State assignment to educational and research institutions (1451).

**Acknowledgment.** A.V.P. thanks Krents A.A. for fruitful discussions and valuable advices regarding the paper findings.

### References

1. I. Babushkin, U. Bandelow, A. G. Vladimirov, "Rotational symmetry breaking in small-area circular VCSELs," Opt. Commun. **284**, 1299-1302 (2011).
2. F. Riechert, G. Verschaffelt, M. Peeters, G. Bastian, U. Lemmer, I. Fischer, "Speckle characteristics of a broad-area VCSEL in the incoherent emission regime," Opt. Commun. **281**, 4424-4431 (2008).
3. S. P. Hegarty, G. Huyet, J. G. McInerney, K. D. Choquette, "Pattern formation in the transverse section of a laser with a large Fresnel number," Phys. Rev. Lett. **82**, 1434-1437 (1999).
4. M. Schulz-Ruhtenberg, I. V. Babushkin, N. A. Loiko, "Transverse patterns and length-scale selection in vertical-cavity surface-emitting lasers with a large square aperture," Appl. Phys. B **81**, 945-953 (2005).
5. J. R. Marciante, G. P. Agrawal, "Spatio-temporal characteristics of filamentation in broad-area semiconductor lasers," IEEE J. Quant. Electron. **33**, 1174-1179 (1997).
6. F. Prati, L. Columbo, "Long-wavelength instability in broad-area semiconductor lasers," Phys. Rev. A **75**, 053811 (2007).
7. V. I. Bespalov, V. I. Talanov, "Filamentary structure of light beams in nonlinear liquids," JETP Lett. **3**, 307-309 (1966).
8. P. K. Jacobsen, J. V. Moloney, A. C. Newell, R. Indik, "Space-time dynamics of wide-gain-section lasers," Phys. Rev. A **45**, 8129-8137 (1992).
9. A. P. Zaikin, N. E. Molevich, "Effect of the cross-relaxation rate on the transverse radiation dynamics of a wide-aperture laser," Quant. Electron. **34**, 731-735 (2004).
10. A. V. Pakhomov, N. E. Molevich, A. A. Krents, D. A. Anchikov, "Intrinsic performance-limiting instabilities in two-level class-B broad-area lasers," Opt. Commun. **372**, 14-21 (2016).

11. C. Sailliot, V. Voignier, G. Huyet, "Filamentation in broad area quantum dot semiconductor lasers," Opt. Commun. **212**, 353-357 (2002).
12. E. Cabrera, O. G. Calderón, S. Melle, J. M. Guerra, "Development of spatial turbulence from boundary-controlled patterns in class-B lasers," Phys. Rev. A **73**, 053820 (2006).
13. E. Cabrera, S. Melle, O. G. Calderón, J. M. Guerra, "Dynamic transition from modelike patterns to turbulentlike patterns in a broad-area Nd:YAG laser," Opt. Lett. **31**, 1067-1069 (2006).
14. S. K. Mandre, W. Elsäßer, I. Fischer, M. Peeters, G. Verschaffelt, "Evolution from modal to spatially incoherent emission of a broad-area VCSEL," Opt. Express **16**, 4452-4464 (2008).
15. D. Amroun, H. Leblond, M. Brunel, C. Letellier, F. Sanchez, "Stabilization of space-time laser instability through the finite transverse extension of pumping," J. Opt. A: Pure Appl. Opt. **10**, 095101 (2008).
16. E. Cabrera-Granado, M. O. Soler Rus, J. M. Guerra, "The influence of the inhomogeneous gain profile on the spatio-temporal dynamics of broad-area class B lasers," Journal of Optics **12**, 035201 (2010).
17. E. Cabrera-Granado, R. Aguade, J. M. Guerra, "Suppression of order–disorder transition in class B lasers due to inhomogeneously broadened gain with fast cross-relaxation rate," J. Opt. A: Pure Appl. Opt. **11**, 045204 (2009).
18. J. Mukherjee, J. G. McInerney, "Spatial mode dynamics in wide-aperture quantum-dot lasers," Phys. Rev. A **79**, 053813 (2009).
19. K. Otsuka, P. Mandel, S. Bielawski, D. Derozier, P. Glorieux, "Alternate time scale in multimode lasers," Phys. Rev. A **46**, 1692-1695 (1992).
20. P. Mandel, M. Georgiou, K. Otsuka, D. Pieroux, "Transient and modulation dynamics of a multimode Fabry-Pérot laser," Opt. Commun. **100**, 341-350 (1993).
21. K. Otsuka, S.-L. Hwong, B. A. Nguyen, "Intrinsic instability and locking of pulsation frequencies in free-running two-mode class-B lasers," Phys. Rev. A **61**, 538151-538156 (2000).
22. K. Staliunas, M. F. H. Tarroja, C. O. Weiss, "Transverse mode locking, antilocking and self-induced dynamics of class-B lasers," Opt. Commun. **102**, 69-75 (1993).
23. F. Prati, L. Zucchetti, G. Molteni, "Rotating patterns in class-B lasers with cylindrical symmetry," Phys. Rev. A **51**, 4093-4108 (1995).
24. A. G. Vladimirov, D. V. Skryabin, "Dynamic instabilities in the interaction of transverse modes in a class-B laser," Quant. Electron. **27**, 887-891 (1997).
25. K. Staliunas, V. J. Sanchez-Morsillo, *Transverse patterns in nonlinear optical resonators* (Berlin, Springer Verlag, 2003).
26. R. Herrero, M. Botey, M. Radziunas, and K. Staliunas, "Beam shaping in spatially modulated broad-area semiconductor amplifiers," Opt. Lett. **37**, 5253-5255 (2012).
27. M. Radziunas, M. Botey, R. Herrero, and K. Staliunas, "Intrinsic beam shaping mechanism in spatially modulated broad area semiconductor amplifiers," Appl. Phys. Lett. **103**, 132101 (2013).
28. S. Kumar, R. Herrero, M. Botey, K. Staliunas, "Suppression of modulation instability in broad area semiconductor amplifiers," Opt. Lett. **39**, 5598-5601 (2014).
29. M. Radziunas, R. Herrero, M. Botey, and K. Staliunas, "Far-field narrowing in spatially modulated broad-area edge-emitting semiconductor amplifiers," J. Opt. Soc. Am. B **32**, 993-1000 (2015).
30. W. W. Ahmed, S. Kumar, R. Herrero, M. Botey, M. Radziunas, and K. Staliunas, "Stabilization of flat-mirror vertical-external-cavity surface-emitting lasers by spatiotemporal modulation of the pump profile," Phys. Rev. A **92**, 043829 (2015).
31. W. W. Ahmed, S. Kumar, R. Herrero, M. Botey, M. Radziunas, and K. Staliunas, "Suppression of modulation instability in pump modulated flat-mirror VECSELs," Proc. SPIE **9894**, 989406 (2016).
32. D. Gailevicius, V. Koliadenko, V. Purlys, M. Peckus, V. Taranenko and K. Staliunas, "Photonic Crystal Microchip Laser," Sci. Rep. **6**, 34173 (2016).
33. M. Eslami, R. Kheradmand, K. M. Aghdami, "Complex behavior of vertical-cavity surface-emitting lasers with optical injection," Phys. Scr. **T157**, 014038 (2013).
34. A. Werner, O. A. Egorov, F. Lederer, "Exciton-polariton patterns in coherently pumped semiconductor microcavities," Phys. Rev. B **89**, 245307 (2014).
35. G. J. De Valcárcel, M. Martínez-Quesada, K. Staliunas, "Phase-bistable pattern formation in oscillatory systems via rocking: application to nonlinear optical systems," Phil. Trans. R. Soc. A **372**, 20140008 (2014).
36. N. Rebrova, T. Habruseva, G. Huyet, S. Hegarty, "Stabilization of a passively mode-locked laser by continuous wave optical injection," Appl. Phys. Lett. **97**, 101105 (2010).
37. N. Rebrova, G. Huyet, D. Rachinskii, and A. G. Vladimirov, "Optically injected mode-locked laser," Phys. Rev. E **83**, 066202 (2011).
38. T. Habruseva, G. Huyet, S. Hegarty, "Dynamics of quantum-dot mode-locked lasers with optical injection," IEEE J. Sel. Top. Quant. Electron. **17**, 1272-1279 (2011).
39. R. M. Arkhipov, T. Habruseva, A. Pimenov, M. Radziunas, S. P. Hegarty, G. Huyet, and A. G. Vladimirov, "Semiconductor mode-locked lasers with coherent dual-mode optical injection: simulations, analysis, and experiment," J. Opt. Soc. Am. B **33**, 351-359 (2016).
40. L. Drzewietzki, S. Breuer, W. Elsäßer, "Timing jitter reduction of passively mode-locked semiconductor lasers by self- and external-injection: Numerical description and experiments," Opt. Express **21**, 16142-16161 (2013).
41. S. Takimoto, T. Tachikawa, R. Shogenji, J. Ohtsubo, "Control of spatio-temporal dynamics of broad-area semiconductor lasers by strong optical injection," IEEE Phot. Tech. Lett. **21**, 1051-1053 (2009).
42. L. A. Lugiato, G.-L. Oppo, J. R. Tredicce, L. M. Narducci, M. A. Pernigo, "Instabilities and spatial complexity in a laser," J. Opt. Soc. Am. B **7**, 1019-1033 (1990).
43. S. Wieczorek, B. Krauskopf, T. B. Simpson, D. Lenstra, "The dynamical complexity of optically injected semiconductor lasers," Physics Reports **416**, 1-128 (2005).